\documentclass[lettersize,journal]{IEEEtran}
\usepackage{amsmath,amsfonts}
\usepackage{algorithmic}
\usepackage{algorithm}
\usepackage{array}
\usepackage[caption=false,font=normalsize,labelfont=sf,textfont=sf]{subfig}
\usepackage{textcomp}
\usepackage{stfloats}
\usepackage{url}
\usepackage{verbatim}
\usepackage{graphicx}
\usepackage{cite}
\usepackage{xcolor}
\usepackage{tikz}    
\begin{document}

\title{Enabling Communication and Control Co-Design in 6G Networks}

\author{Onur Ayan, Nikolaos Pappas, Miguel Angel Gutierrez Estevez, Xueli An,  Wolfgang Kellerer
\thanks{O. Ayan, M. A. Gutierrez Estevez, and X. An are with Huawei Technologies, Munich, Germany. \\ (e-mail: \{onur.ayan, m.gutierrez.estevez, xueli.an\}@huawei.com)\\ W. Kellerer is with the Chair of Communication Networks, Technical University of Munich, Munich, Germany \\ (e-mail: wolfgang.kellerer@tum.de).\\ N. Pappas is with Linköping University, Linköping, Sweden \\ (e-mail: nikolaos.pappas@liu.se).}
}

\maketitle

\begin{abstract}
Networked control systems (NCSs), which are feedback control loops closed over a communication network, have been a popular research topic over the past decades. Numerous works in the literature propose novel algorithms and protocols with joint consideration of communication and control. However, the vast majority of the recent research results, which have shown remarkable performance improvements if a cross-layer methodology is followed, have not been widely adopted by the industry. In this work, we review the shortcomings of today's mobile networks that render cross-layer solutions, such as semantic and goal-oriented communications, very challenging in practice. To tackle this, we propose a new framework for 6G user plane design that simplifies the adoption of recent research results in networked control, thereby facilitating the joint communication and control design in next-generation mobile networks.
\end{abstract}


\section{Introduction}
\label{sec:introduction}
Networked control systems (NCSs) are feedback control loops that are closed over a communication network. An NCS comprises one or more sensors, actuators, and a controller regularly exchanging information to monitor and control a physical process. Factory automation, mobile robots, and unmanned aerial vehicles (UAVs) are only a few of the most prominent examples of NCSs, which belong to the key use cases of the sixth generation (6G) mobile networks \cite{giordani2020toward}.

Joint communication and control has been a popular research topic over the past decades \cite{walsh2001scheduling, scheuvens2021state, han2022flexible}. Numerous scientific papers have been published that propose novel solutions to mitigate the adverse effects of the network on the control system's performance. Most of these works, particularly those focusing on control-aware networking, assume an intertwined operation of communications and control. That is, the network is considered to be aware of the control application's characteristics (i.e., system dynamics, sampling frequency, control law) and the content of data from the user plane that are being transmitted.

Contrarily, if we look at the core principles of today's mobile networks, we find fundamental differences between the approach taken by the research and standardization communities. Since the inclusion of machine-type communications, Internet of Things (IoT), and time-critical applications as key use cases of cellular networks, starting from the fourth generation (4G), most of the standardization effort has flown into offering an extremely reliable and fast service to verticals. Particularly, in an industrial ecosystem consisting of multiple heterogeneous NCSs, the communication network is expected to transmit every generated packet by a source to the destination as if they were connected by an ``almost ideal" link\footnote{Analogy of a ``data pipe" has been used in the literature\cite{kountouris2021semantics}.}. This led to the popularity of ultra-reliable low latency communications (URLLC) in 5G to support time-critical applications, which require multiple nines of reliability and end-to-end latency at the millisecond level. For instance, \cite{3gppts22104} defines a minimum reliability target of $99.9999\%$ and a maximum end-to-end latency less than a sampling period (i.e., transfer interval) and zero survival time for process automation and closed-loop control. 

Although it is ideal to interconnect the components of a feedback control loop via perfect links, this is not always mandatory to get a close-to-optimal performance for many applications. An existing work supporting this argument is \cite{mager2019feedback}, which conducts a case study involving an inverted pendulum, a prevalent application in control theory textbooks due to its open-loop unstable nature and fast dynamics. In their work, the authors show that despite the gradual performance degradation for increasing packet loss rate, the control system's stability is preserved up to a packet error rate of 75\%. 
This raises the question of whether URLLC is a mandatory enabler of industrial applications. As stated in \cite{scheuvens2021state} and \cite{han2022flexible}, enforcing such stringent requirements on the network suffers from over-provisioning in bandwidth and energy. As a result, the network resource utilization could be more efficient, which may prevent the admission of new flows due to resource scarcity.

In contrast to 5G, 6G systems are envisioned to achieve a full convergence of communication and control \cite{saad2020vision}. However, this mandates specific end-to-end system architecture and protocol design modifications. In this work, we discuss the main shortcomings of 5G that we were able to identify from the perspective of recent research activities on NCSs and propose a new framework facilitating the communication and control co-design in the next-generation mobile networks. Our framework builds on and extends the core concepts of the \textit{Open Platform Communication Unified Architecture (OPC UA)}, which is an industrial standard developed for data exchange between devices such as sensors and controllers to increase interoperability \cite{opc2022part1}

The remainder of this work is organized as follows. Section \ref{sec:background} provides a background on various relevant topics from cellular and industrial networking domains. In addition, we elaborate on the key design principles in 5G hindering the communication and control co-design. In section \ref{sec:framework}, we discuss how this can be overcome with the help of our proposed framework and explain its key distinctive features. Section \ref{sec:numerical_evaluation} presents a numerical evaluation of our proposed framework based on a simulation-based case study.
\section{Existing Approaches and Limitations}
\label{sec:background}

\subsection{5G User Plane Architecture}
\label{subsec:5G_up_arch}
5G network architecture assumes the separation of user and control planes. The \textit{control plane (CP)} carries out network-internal tasks such as network configuration, connection establishment, and resource management. Transmission of the application data is handled by the \textit{user plane (UP)}, although the name may be misleading for control systems society. The UP is responsible for forwarding real-time application data from one network node to another.

\begin{figure}
	\centering
	\includegraphics[width=.9\columnwidth]{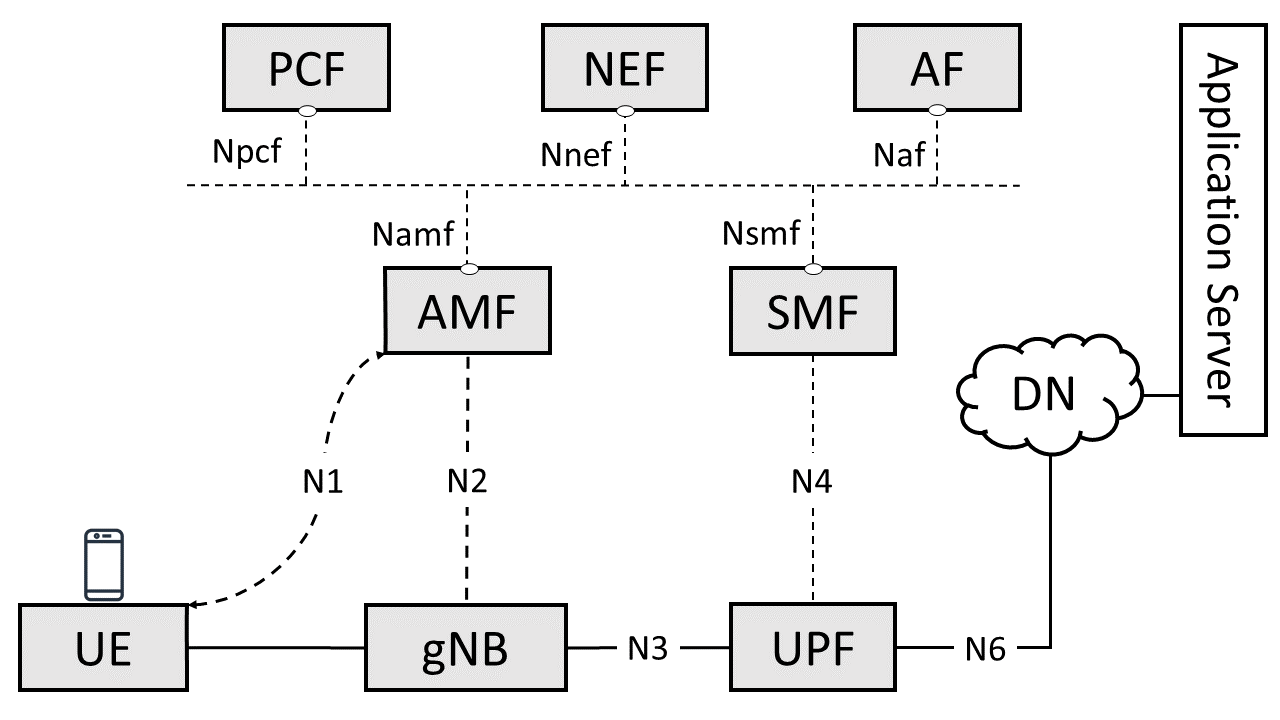}
	\caption{A simplified architecture of 5G radio access network (RAN) and core network (CN). The user plane data from a UE to the application server is transmitted first on the uplink to gNB, from which it they are forwarded to the data network (DN) via the serving user plane function (UPF). Other important CN functions are policy control function (PCF), network exposure function (NEF), application function (AF), access and mobility management function  (AMF), and session management function (SMF). We refer to \cite{3gppts23501} for further details.}
    \label{fig:5G_architecture}
\end{figure}

A real-time control application closed over a 5G network generates data through IP packets or Ethernet frames. For instance, if a sensor application runs on a 5G capable mobile device, i.e., user equipment (UE), whereas the controller is deployed on a remote application server, the data packets carrying the sensor measurements need to be transmitted first to the 5G base station called gNodeB (gNB) using the 5G new radio (5G-NR) air interface. Following a successful reception, the gNB forwards the data to another node called the user plane function (UPF), from where they are forwarded to the application server hosting the remote controller process, typically located behind a data network (DN). Fig. \ref{fig:5G_architecture} depicts the simplified 5G radio access- and core network architecture. In 5G, every connection between a UE and the 5G system is established between the UE and a UPF. In 3GPP terms, this connection is called a \textit{PDU session}. 5G supports IP and Ethernet-type PDU sessions depending on the deployment scenario and use case. 

\subsection{Quality of Service (QoS) Management in 5G}
\label{subsec:5G_qos}
According to the latest specifications, the \textit{QoS flow} is the finest granularity of QoS differentiation in a PDU session\cite{3gppts23501}. A QoS flow may either be of type \textit{guaranteed bit rate (GBR)} or a non-GBR, while the GBR is typically applicable to NCSs. Without going into much detail, each QoS flow is mapped to specific QoS parameters that characterize the service it demands from the 5G System (5GS). These parameters can be in the form of maximum packet loss rate, packet delay budget between a UE and UPF, and guaranteed or maximum flow bit rate. The specifications define parameters applicable to industrial applications, such as nominal message size, transfer interval, survival time, and maximum end-to-end latency. According to \cite{3gppts22104}, a typical closed-loop control application sends twenty bytes of data every ten milliseconds, whereas the maximum end-to-end latency is given as one transfer interval, i.e., $10$ ms. It is of utmost importance to emphasize that each packet belonging to a QoS flow is characterized by a single QoS parameter set, hence treated equally by the 5G's UP protocol stack, which may not be optimal in wireless NCSs.

\subsection{Communication Models Employed in Industrial Networks}
\label{subsec:communication_models}
An industrial application typically involves various sensors and controllers generating periodic or sporadic (i.e., time-triggered or event-triggered) messages that a 5G system must serve. Such systems follow a \textit{push-based} communication model, in which an information source ``pushes'' data packets toward the destination, meaning the source is the entity deciding for or against a transmission. 

In industrial networks, \textit{publish-subscribe (PubSub)} paradigm stands out as a sub-category of push-based systems. In PubSub, a ``publisher'' sends the produced data to a message-oriented middleware from which they are forwarded to subscribers\footnote{A message-oriented middleware is also known as a ``broker'' in the literature.}. The distinctive feature of PubSub is to decouple publishers from subscribers, i.e., the subscribers and publishers do not know about each other because of the intermediate middleware. The PubSub pattern is already used for real-time data exchange in the industrial standard OPC UA. The well-known robot operating system (ROS) is another example of the adoption of the PubSub model. 

As an alternative to push-based systems, numerous existing works adopt a \textit{pull-based} communication model, in which packets are sent only when the information consumers, typically controllers, actuators, and supervisory computers, request it from the information producers. This not only introduces additional delay into the feedback control loop, as the pull request must be transmitted preceding the actual transmission, but also, increases the communication overhead. Despite its disadvantages, the pull-based model helps the mobile network to potentially free communication and computation resources, particularly in large-scale networks, by limiting the exchanged data to only those requested by the data consumer, thereby allowing more important and urgent transmissions to capture the limited resources more efficiently.

It is noteworthy that the control-aware resource management for NCSs, which focuses on granting network resources to those control applications that are estimated to have the most important information to transmit, contains some characteristics from the push-based and pull-based communication models \cite{scheuvens2021state, uysal2022semantic, walsh2001scheduling, ayan2023optimal}\footnote{Note that scheduling is a data-link layer problem in its essence to orchestrate the access to the shared medium.}. Although it is not the receiving application directly requesting particular information from the data source, this responsibility is offloaded to the scheduler, which is guided by exposing specific characteristics of the system dynamics or control task. On the sender side, the source publishes data periodically that is admitted into the network, hence resembling the push-based model. In essence, this is nothing other than migrating the pull-based communication between two application end-points down to the access stratum, while the sending application is only responsible for keeping the data fresh on either side of the air interface. Such an approach is very effective when it comes to improving QoC while at the same time being more delay-friendly than the vanilla pull-based model.

\subsection{Shortcomings of Today's Cellular Networks}
\label{subsec:missing_blocks}
\textbf{Lack of a ``Quality of Control'' framework:} NCSs and many other example vertical applications exchange information to execute a particular task in the physical world. For instance, a sensor sends real-time process data to a controller, through which the system state is driven to the desired value (setpoint). The application's performance is quantified neither by the amount of data packets that are successfully sent by the network nor by how fast they arrive at the destination. Instead, the performance is entirely characterized by the success level in the (control) task achievement, which is also referred to as \textit{Quality of Control (QoC)} in the literature. 

In contrast to industrial use cases, the current specifications include a set of considerations on Quality of Experience (QoE), which capture the application layer performance for multimedia applications, such as real-time audio and video streaming as well as virtual reality and are reported to the 5GS \cite{3gppts26247, 3gppts26118}. Examples are ``representation switch events'', ``playout delay for media start-up'' and ``buffer level''.

In contrast to the QoE framework, 5GS does not feature any KPIs that quantify the QoC. Instead, the packet error rate, average bit rate, end-to-end latency, and round-trip time are used as service requirements that the 5GS has to meet, and the actual application performance, given the control task at hand, is not considered. Some prominent examples from the existing literature that are used to quantify QoC are, including but not limited to, linear-quadratic-Gaussian (LQG) cost, mean squared error (MSE), cost of actuation error, and integral absolute error (IAE).

\textbf{A semantics-aware user plane design:} As of today, 5G support for industrial applications follows an information-agnostic design principle, meaning that neither the content of the incoming data packets nor their attributes, such as freshness, importance, and accuracy are taken into account. Various works in the literature discuss the key principles and benefits of semantics-aware networking \cite{kountouris2021semantics, uysal2022semantic}. In the context of networked control systems, introducing semantic awareness to the 5GS helps the network utilize the limited resources more efficiently and improve the QoC by prioritizing more important packets.

The semantics-aware protocol design has played a central role in the NCS research conducted over many decades. A typical example is the \textit{event-triggering (ET)} technique, which allows the transmission of a data packet only if the content is classified as ``important'' according to a threshold-based policy. In addition to reducing the amount of admitted data, ET is also used to reduce end-to-end latency and packet loss due to congestion in the network. Another well-known example is using \textit{network-induced error} for user prioritization to improve the QoC \cite{walsh2001scheduling, ayan2023optimal}.

\section{A Goal-Oriented Framework for NCSs in 6G}
\label{sec:framework}


\subsection{The Monitored Data Unit}
\label{subsec:mdu}
Industrial use cases typically involve regular and frequent transmission of real-time process data that live in the physical world, such as temperature, pressure, and position. As also briefly mentioned in section \ref{subsec:5G_up_arch}, these data are typically formed either into IP packets or Ethernet frames and subsequently forwarded to lower layers of the 5G communication stack. 

However, this approach has one major fundamental issue that renders the communication and control co-design challenging: From the network's perspective, the payload section of an IP packet is nothing more than a set of bytes, as no information about its structure is revealed to the 5GS. Consider the following simple example: suppose an IPv4 packet with a $32$ bytes payload will be transmitted on the uplink (UL). In this case, the 5GS does not know whether it is a concatenation of multiple system states, e.g., temperature reading and position information, or a single sensor measurement that is 32 bytes long. Consequently, it does not know whether two subsequent IP packets contain measurements of the same quantities at different points in time or whether their content is entirely independent, e.g., the earlier packet carrying position information and the latter velocity. In fact, according to the OPC UA specifications, a data packet that a sensor or controller sends can contain the concatenation of multiple ``pieces of information'', e.g., sensor readings and actuation signals. Thus, two consecutive packets are not guaranteed to contain the same information set \cite{opc2022part14}. 

\begin{figure}
	\centering
	\includegraphics[width=\columnwidth]{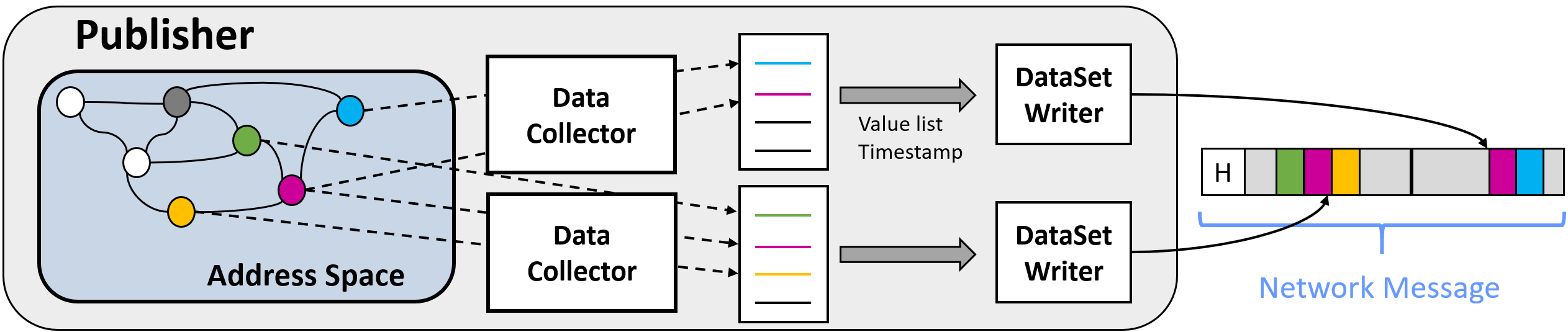}
	\caption{Information flow from the perspective of a \textit{Publisher} in OPC UA. The publisher maintains an \textit{information model} consisting of an \textit{Address Space} of multiple \textit{nodes}. A \textit{node} corresponds to a real object whose attributes are monitored. The figure is recreated from the OPC UA specifications. 
 }
    \label{fig:publisheddataset}
\end{figure}

This is illustrated in Fig. \ref{fig:publisheddataset}, in which a \textit{publisher} sends real-time data as \textit{network message}s\footnote{Here, we use italic font to emphasize OPC UA terms.}. For each published item, few parameters dictate how and when a new update is transmitted. For instance, for \textit{cyclic dataset}s, monitored data are sent only once every \textit{publishing interval}, corresponding to time-triggered sampling in control theory. Contrarily, \textit{acyclic dataset}s are transmitted after passing a \textit{filter} that defines the data reporting condition\footnote{The \textit{filter} fulfills the same role as \textit{event-triggering}.}. If \textit{filter}s are employed, two consecutive \textit{network message}s may not contain the same dataset. As a result, if the underlying communication system discards an outdated packet to replace it with a fresher one, a common approach in the existing literature that has proven efficient in increasing information freshness, it may discard certain content not included in the new packet.

The issue mentioned above calls for \textit{atomization} of messages into finer pieces that cannot be further fragmented without losing their context. To address this, we define a \textit{monitored data unit (MDU)} as the timestamped digital representation of a physical quantity that is transmitted from a source application to one or more receiving applications and cannot be divided further into smaller chunks before being processed and used on the receiving end. For instance, a temperature sensor reading of four bytes is an example of an MDU, which requires to be received by the monitor in its entirety for correct interpretation.

\subsection{6G RAN as a Goal-Oriented Middleware}
\label{subsec:broker}

As we have established in section \ref{subsec:missing_blocks}, the communication purpose in NCSs is the execution of certain control tasks. 

However, the aforementioned design principle separating the network from applications does not leave much room for cross-layer considerations, which limits the 5G network's role to a data- and goal-agnostic data delivery system. As a result, to leverage the recent research on goal-oriented communications for NCSs, 6G RAN architecture must undergo a few modifications, which we aim to tackle with a new UP framework.

As an essential building block of our framework, we bring the notion of an MDU into 6G's scope. In particular, we enable the 6GS to logically relate two distinct measurements of a process, e.g., two consecutive IP packets that are 10 milliseconds apart. This gives the 6GS the ability to treat each MDU differently, for instance, by storing these in MDU-specific transmission buffers throughout their service and applying differentiated treatment with packet-level granularity, which is a central idea in semantic communications. The two most obvious ways to achieve this in practice are either providing each MDU as a separate (QoS) flow or in a single flow but with unencrypted metadata such as a flow label\footnote{Note that this does not imply any access to the payload by the network, therewith, respects the end-to-end data privacy.}. For instance, data from an IMU sensor could be assigned a unique label that allows for their identification by the 6GS.

Having introduced how the real-time data is provided to the 6G system, we now characterize the treatment of MDUs by the UP protocol stack. To that end, we introduce a new layer called \textit{semantic aggregation layer (SAL)} that receives and identifies MDUs as described above. Within the SAL, we define four main sub-functions:

\begin{enumerate}
    \item The \textit{data handler (DH)} serves as the entry point of MDUs to the SAL. DH is responsible for the identification of MDUs upon arrival and their storage afterward. The DH is also the entity that is responsible for filtering and prioritizing MDUs based on a pre-configured policy. An example is the prioritization of MDUs with the highest importance, which may be provided by the application as readable metadata per MDU to assist the DH. Another example is the consideration of data freshness captured by the age of information (AoI) metric or the ability to discard PDUs actively based on their importance or staleness. 
    \item The \textit{data writer (DW)} is the function responsible for composing the outgoing data, i.e., the SAL PDU, which is retrieved from the DH. Each SAL PDU contains all the necessary information to enable the receiving SAL entity to decompose it into the MDUs contained within.
    \item The \textit{data reader (DR)} is the counterpart of the DW on the receiving side, which decomposes a successfully received PDU into one or more MDUs by their IDs. Additionally, it is responsible for sending MDU ACKs to the transmitting SAL upon their successful reception. 
    \item The \textit{session handler (SH)} manages high-level tasks such as (de-)registration of an MDU label, ID assignment, and prioritization rule configuration.
\end{enumerate}

The SH and DH at the receiving entity play an essential role in the operation of our proposed framework. When a data producer or consumer desires to send or receive a particular MDU, respectively, the receiving SH is informed about the MDU label, source, and destination addresses\footnote{We would like to mention that there is an MDU label alignment step happening in the application layer, that is external to the 6G network.}. In other words, the information consumer(s) ``subscribes'' to a particular MDU that it is interested in receiving from a specific producer, i.e., ``publisher''. Accordingly, the receiving DH composes outgoing UP packets (IP packets or Ethernet frames depending on the session type) based on the subscription information through flow IDs or labels\footnote{Note that the SH needs to configure the DH accordingly so that the receiving DH can compose the outgoing data packets with the correct source and destination addresses/ports.}. This makes the 6G RAN operate as a message-oriented middleware of a publish/subscribe (PubSub) model. Last but not least, the receiving SAL RAN entity is capable of requesting MDUs by label/ID from the transmitting SAL entity. The main benefits of the newly proposed framework are discussed in the next section.

\begin{figure}
	\centering
	\includegraphics[width=\columnwidth]{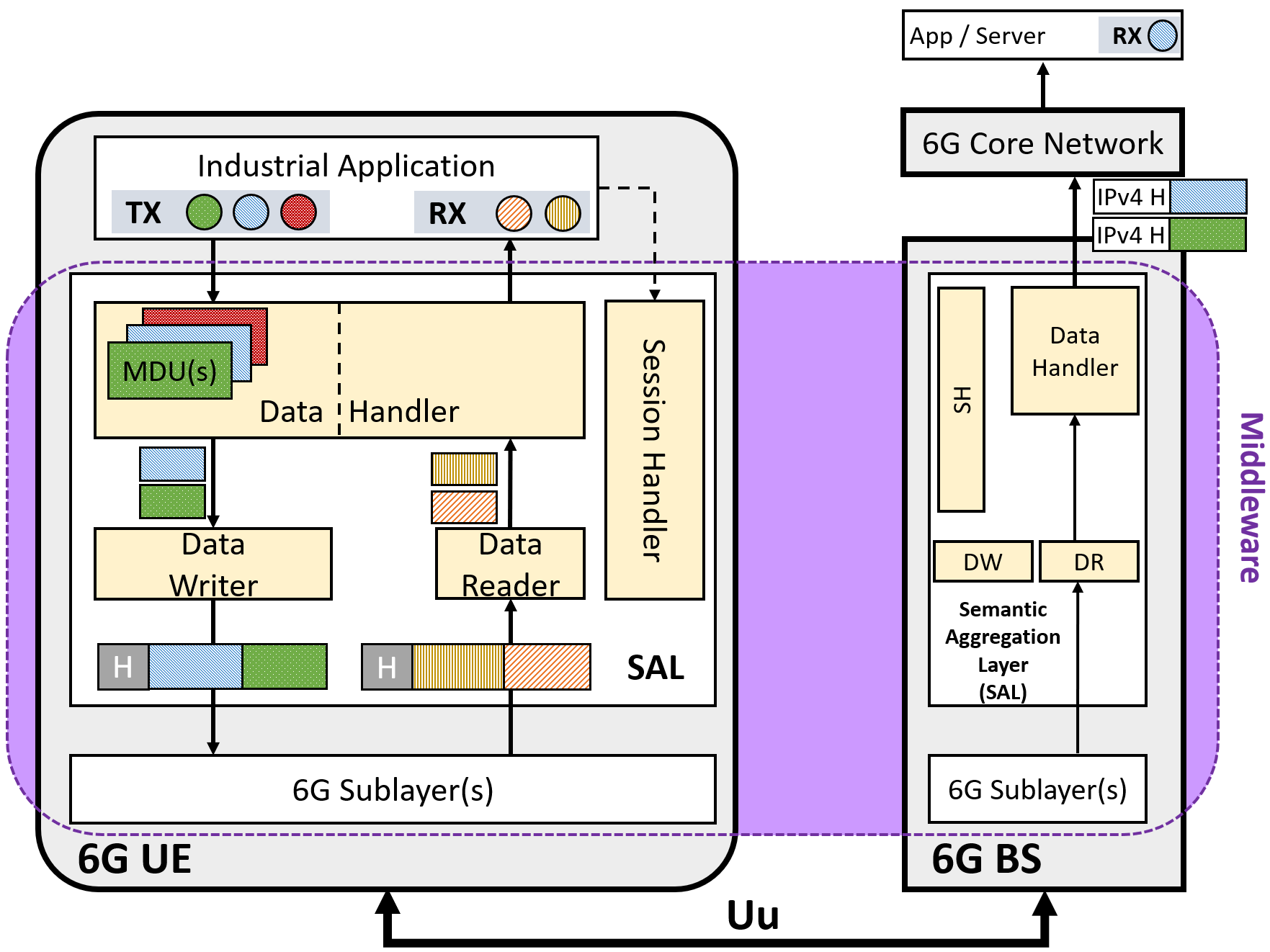}
	\caption{6G RAN user plane serving as a semantics-aware middleware for time-sensitive industrial applications.}
    \label{fig:framework}
\end{figure}

\subsection{Relationship to Communication and Control Co-design}
\label{subsec:benefits}
The SAL we have introduced in section \ref{subsec:broker} may appear as an ordinary (sub-)layer of 5G RAN that first forwards data packets after some internal processing and header encapsulation to lower layers. However, there are fundamental functional characteristics distinguishing it from such. 

Firstly, the SAL decides whether and when a particular content is transmitted, typically a role played by the application layer. This calls for a new set of service agreements and KPIs between the control application and the communication system, such as the maximum allowed time between two successful MDU transmissions, a metric strongly affecting the peak (maximum) AoI at the subscribers. Note that this is different from the push-based systems, where the communication system must meet pre-defined service requirements on exogenous data arrivals because the communication network now can determine the exact transmission time of an MDU without causing additional staleness during service.

Secondly, the event-triggering mechanism, which has been a binary decision made in the application layer for or against a transmission, is now enhanced with a third option: ``transmit if space''. That is, an MDU, which the application may have discarded due to low importance, can now be accommodated in an uplink transmission instead of the padding bits. This increases the network's spectral efficiency and postpones the occurrence of the next ``event'' to a later point in time, thereby improving the QoC.

An important research problem in NCSs is the dynamic distribution of the limited network resources among multiple applications, especially when these are of heterogeneous type. Since the radio link between the UEs and the BS has been considered the main bottleneck in cellular remote monitoring and control use cases, the 6G RAN should be capable of prioritizing the most important transmissions in uplink and downlink directions. It is essential to emphasize that this is not equivalent to granting users time and frequency resources without specifying how these should be utilized. On the contrary, the research has shown significant gains in QoC if the RAN can assess semantics and context and consequently specify \textit{for which particular information the allocated resources should be used}. 

Consider a UL transmission of an MDU to demonstrate how this can be achieved by our framework. As the receiving SAL is now able to identify and differentiate MDUs, it can also request specific information by signaling its (MDU) ID to the transmitting SAL. Thus, the UL radio resources can be used more efficiently when compared to the case where UE selects the next PDU in the transmission buffer in an information-agnostic fashion. As a result, the 6G RAN, which connects the publishers of MDUs to their subscribers as illustrated in Fig. \ref{fig:framework}, operates as a \textit{goal-oriented} middleware that takes the control-dependent semantics of information into account, therewith, achieving a full convergence of control and communication.

\section{Performance Evaluation}
\label{sec:numerical_evaluation}
We consider a simulation-based case study with a single transmitter and $N$ receivers. The transmitter publishes periodic data of $N$ independent physical processes and admits the generated data packets into a communication network. The network is configured with the subscription information and is responsible for delivering the data traffic to the corresponding subscribers. The scenario corresponds to $N$ decoupled NCSs executing parallel control tasks in the physical world through the provision of real-time data\footnote{As an example, consider a video camera transmitting the positions of $N$ physical objects to robot controllers for task execution.}.

We consider four alternatives for the composition of published data:  
\begin{itemize}
    \item \textbf{Unfiltered compound (UC):} The UC strategy characterizes a publisher that transmits its entire sampled dataset without any filtering mechanism (e.g., event-triggering) in a single packet of constant size.
    \item \textbf{Filtered compound (FC):} According to the FC strategy, the publisher uses an event criterion for each MDU to decide whether it should be included in the next packet transmission.
    \item \textbf{Unfiltered Atomic (UA):} Here, the publisher sends sensor measurements without any filtering mechanism. In contrast to UC, each MDU is sent in a different data packet.
    \item \textbf{Filtered Atomic (FA):} As the name suggests, this combines the UA strategy and event-triggering for each MDU.
 \end{itemize}

The UC and FC strategies result in a single QoS flow from the network's perspective. Contrarily, both UA and FA allow the publisher to send each sensor measurement as a separate flow, e.g., via separate UDP ports. This preserves the MDU granularity, which exists in the application layer and also in the network throughout service. 

We utilize non-linear functions of AoI per MDU that depend on the control system parameters\footnote{An example of control-dependent non-linear functions of AoI can be found in \cite{ayan2023optimal}.} for the joint control- and communication flow prioritization by the DH. For simplicity, we model the UL as a packet erasure channel with a loss probability of $10\%$. We implement the ET mechanism (applicable to FC and FA) as in the OPC UA specifications\footnote{The ET mechanism is referred to as \textit{Deadband Filter} in the specifications.}.

\begin{figure}
	\centering
	\includegraphics[width=\columnwidth]{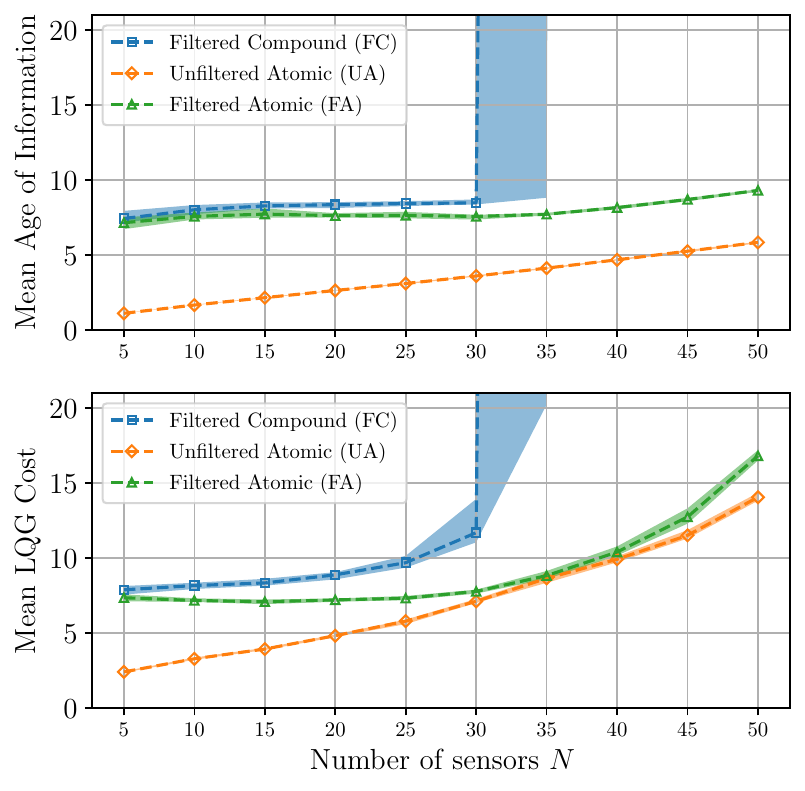}
	\caption{Mean Age of Information (AoI) and linear-quadratic-gaussian (LQG) cost in the network for varying number of NCS, i.e., $N$. A lower AoI and LQG indicates a higher performance w.r.t. the measured metric. Semi-transparent colored regions depict the minimum and maximum mean values throughout our simulations comprising twenty repetitions.}
    \label{fig:results}
\end{figure}

We present Fig. \ref{fig:results} containing the main results of our study, namely, the network's information freshness and control performance captured by the mean AoI and LQG cost, respectively. We do not include UC in the figure for presentation purposes. Nevertheless, we would like to mention that it performs significantly worse than the other strategies, which is expected due to its semantics-agnostic nature. Fig. \ref{fig:results} indicates that the FC compound mechanism is outperformed by both UA and FA policies, not only for a higher number of NCSs but also for the lowest considered one, i.e., $N = 5$. Surprisingly, UA performs the best among all, delivering a higher control performance than FA. For a lower number of sensors, this can be explained by the under-utilization of available network resources caused by the event triggering in the application layer. In other words, although the UE can accommodate more bits in an uplink transmission (in the transport block), it fails to do so as certain data is filtered in by the application, eventually resulting in \textit{padding bits}. More interestingly, the FA is behind UA also for larger $N$ due to inadequate information available to the ET mechanism while determining the packet importance. This is caused by the filter implementation in the OPC UA standard, which admits a packet based on the previous transmitted state. In contrast, it does not know its actual status, i.e., whether it has been transmitted or discarded by the UE but also lost during an UL transmission\footnote{This is an example showing that the OPC UA is mainly designed with a lossless wired connectivity in mind.}. Note that an end-to-end acknowledgment mechanism does not solve this issue as the 5G RAN can discard packets prior to any transmission, e.g., if a \textit{discard timer} of a transmitting PDCP entity is configured.

\section{Conclusion}
\label{sec:conclusion}
Cross-layer design has shown great potential for performance and efficiency maximization in NCSs. Application and network engineers need to depart from the conventional design principles that render cross-layer solutions infeasible and, therefore, exploit the recent research activities on joint communication and control. In this work, we have briefly overviewed key design principles necessary for enabling joint communication and control. We have shown how these can be employed in 6G without violating mobile communications standards' core requirements and capabilities.

\bibliographystyle{IEEEtran}
\bibliography{references}

\vfill

\end{document}